\documentclass[prb,preprint,superscriptaddress,nofootinbib]{revtex4}

\newcommand{\Xe}{$^{129}$Xe}
\newcommand{\F}{$^{19}$F}
\newcommand{\caf}{CaF$_2$}

\usepackage{graphicx}		
\usepackage{dcolumn}		
\usepackage{bm}			
\usepackage{multirow}		
\usepackage{subfig}
\usepackage{caption}				

\usepackage{appendix}
\usepackage{amsmath, amssymb, bbm}
\usepackage{array}
\begin{document}

\title{Phase Relationship Between the Long-time Beats of Free Induction Decays and Spin Echoes in Solids }

\author{E. G. Sorte} \affiliation{Department of Physics, University of Utah, 115 South 1400 East, Salt Lake City, Utah 84112-0830, USA} 
\author{B .V. Fine} \email{B.Fine@thphys.uni-heidelberg.de} \affiliation{Institute for Theoretical Physics, University of Heidelberg, Philosophenweg 19, 69120 Heidelberg, Germany}
\author{B. Saam} \email{saam@physics.utah.edu} \affiliation{Department of Physics, University of Utah, 115 South 1400 East, Salt Lake City, Utah 84112-0830, USA} 

\date{\today}

\begin{abstract}

Recent  theoretical work on the role of microscopic chaos in the dynamics and relaxation of many-body  quantum systems has made several experimentally  confirmed predictions about the systems of interacting nuclear spins in solids, focusing, in particular, on the shapes of spin echo responses measured by nuclear magnetic resonance (NMR).  These predictions  were based on the idea that the transverse nuclear spin decays evolve in a manner governed at long times by the slowest decaying eigenmode of the quantum system, analogous to a chaotic resonance in a classical system. The present paper extends the above investigations both theoretically and experimentally.  On the theoretical side, the notion of chaotic eigenmodes is used to make predictions about the relationships between the long-time oscillation phase of the nuclear free induction decay (FID) and the amplitudes and phases of  spin echoes.  On the experimental side, the above predictions are tested for the nuclear spin decays of \F\ in \caf\ crystals and \Xe\ in frozen xenon. Good agreement between the theory and the experiment is found.    
\end{abstract}

\maketitle

\section{\label{intro} Introduction}

The role of microscopic chaos in the observable behavior of macroscopic objects is a notoriously difficult elusive issue\cite{Gaspard-98,Gaspard-98A,Dettmann-99,Grassberger-99,Gaspard-99}. Exploring it requires joint analytical, numerical and experimental efforts.  The focus of this  paper is on a possible implication of microscopic chaos for free induction decays (FIDs) and spin echoes measured by nuclear magnetic resonance (NMR) in solids.  

Previously, one of us (B.V.F.) has argued~\cite{Fine-00,Fine-03,Fine-04} that, as a consequence of microscopic chaos induced by generic non-linear interaction between nuclear spins, the long-time behavior of nuclear FIDs in solids has the universal long-time form

\begin{equation}
F(t) = A e^{- \gamma t} \hbox{cos}(\omega t - \varphi_a),
\label{ltform}
\end{equation}
where $A$, $\gamma$, $\omega$ and $\varphi_a$ are constants. Depending on the microscopic Hamiltonian of interacting nuclear spins, the frequency $\omega$ may be equal to zero. However, in the most common case of the magnetic dipole interaction, $\omega$ has a finite value (see the discussion in Ref.\cite{Fine-04}). Indeed, such behavior was observed as generic  in the experimental\cite{Engelsberg-74,Morgan-08,Sorte-11,Meier-11} and numerical\cite{Fabricius-97,Fine-03} studies of quantum and classical spin systems. In a typical case, the constants $\gamma$ and $\omega$ fall on the fastest natural microscopic time scale of the nuclear spin system, thereby precluding any explanation of the above behavior in terms of a damped harmonic oscillator --- such an explanation would require the separation of time scales between the slow observable $F(t)$ and much faster microscopic motion. 

The theoretical analysis of Ref.\cite{Fine-04} predicted only the functional form (\ref{ltform}) of the long-time FID behavior without predicting the parameters $A$, $\gamma$, $\omega$ and $\varphi_a$. A later paper~\cite{Fine-05} went further and predicted that different spin echoes initiated in the same system by perturbing the FID with almost any sequence of radio-frequency (rf) pulses would have different initial behavior but then evolve to exhibit the long-time behavior \eqref{ltform} characterized by the same time constants $\gamma$ and $\omega$. This prediction was confirmed experimentally in Refs.\cite{Morgan-08,Sorte-11} for hyperpolarized solid xenon and CaF$_2$. 

The present paper explores the chaos-related notion that the long-time behavior of many-spin density matrices created in the course of the FID has a self-similar form accompanying exponentially decaying oscillations. In Section~\ref{sec:theory}, we show theoretically that for the spin echoes initiated by perturbing the FID in the above long-time regime, all possible shapes of the echo responses are superpositions of two basics shapes with the relative weight of each shape determined by the phase of the FID oscillations at the time of the echo pulse. This two-shape decomposition is predicted to lead to a definite relationship between the long-time oscillation phase of the original FID and the  long-time oscillation phases of the echo responses. In Section~\ref{sec:experimental_procedures}, we verify the above predictions experimentally for \caf, and include measurements on solid xenon in the appendix.

\section{\label{sec:theory} Theory}

The prediction of universal behavior of quantum spin systems in Refs.~\cite{Fine-04,Fine-05} was based on the conjecture that the long-time behavior \eqref{ltform} is a manifestation of the slowest decaying chaotic eigenmode of the time-evolution operator, similar to a Pollicott-Ruelle resonance\cite{Ruelle-86,Gaspard-98}.  Such eigenmodes control not just one observable quantity $F(t)$ but also the evolutions of many-spin density matrices within the system

\begin{equation}
\rho_{kl}(t) = \rho_{0, kl} e^{-(\gamma + i \omega) t} +  \rho_{0, kl}^{\dagger} e^{-(\gamma - i \omega) t} 
\label{rho}
\end{equation}
where $\rho_{kl}(t)$ is the density matrix for any finite subsystem of the entire spin system, i.e. $\rho_{kl}(t)$ can be a one-spin density matrix, two-spin density matrix, or, in general, an $n$-spin density matrix, provided $n$ is much smaller than the total number of spins in the system. As is often done in the NMR literature, Eq.~\eqref{rho} represents the leading correction to the infinite temperature density matrix $\rho_{kl} = {\mathbbm 1}$. The term $\rho_{kl} = {\mathbbm 1} $ does not contribute to the measured spin polarization.  This high-temperature approximation should remain valid as long as the initial nuclear polarization is not too large, in the sense that the initial energy of the nuclear spin system with respect to the effective Hamiltonian of nuclear spin-spin interaction in the Larmor rotating reference frame\cite{Slichter-90} is close to the energy of the infinite temperature state. If the initial polarization is too large, then the system is expected to relax to a finite temperature equilibrium determined by its initial energy, in which case Eq.(\ref{rho}) would represent the correction to the equilibrium density matrix for the final temperature.

Equation (\ref{rho}) is the only connection between the theoretical analysis in this paper and the notion of microscopic chaos. Namely, the assumption of microscopic chaos justifies the proposition that well-defined relaxational eigenmodes of the time-evolution operator of the entire system exist. In turn, the notion of an eigenmode of the time-evolution operator implies that the values of $\gamma$ and $\omega$ do not depend on the order of the density matrix.  Whatever the initial form of the $n$-spin density matrix, the long-time behavior would then be dominated by the slowest chaotic eigenmode of form \eqref{rho} (among those compatible with the symmetry of the initial density matrix). The time-independent non-Hermitian form of $\rho_{0, kl} $ for a given order of the density matrix, as well as the values of $\gamma$ and $\omega$, are determined by the microscopic Hamiltonian of the system.  While the above connection to microscopic chaos is very indirect,  we are not aware of any other framework justifying Eq.(\ref{rho}). Our assumption of microscopic chaos  is, in turn, motivated by the non-integrable character of nuclear dynamics governed by the nuclear spin-spin interaction Hamiltonian in the Larmor rotating reference frame~\cite{Fine-04}.

The experimental evidence available so far is obtained from the total polarization of nuclear spins and, as such, indicates that the decay \eqref{rho} is certainly present in the behavior of the one-spin density matrix, but not necessarily two-spin, three-spin and progressively higher-order spin density matrices. Higher-order density matrices are responsible for higher-order nuclear correlations  (spin coherences in NMR language). The FID starts from a factorizable density matrix for the entire system~\cite{Slichter-90}, meaning that the initial nuclear spin configuration is uncorrelated. Therefore, the expectation behind Eq.~\eqref{rho} is that the higher-order correlations first develop dynamically and then start decaying~\cite{Cho-05}, eventually approaching form~\eqref{rho}. The effect of the echo pulse does not reverse but rather modifies the higher-order correlations. The predictions made below about the relationship between the shape of the echo response and the phase of the long-time FID beats at the time of the echo pulse are expected to be incorrect if the many-spin density matrices preceding the echo pulse do not exhibit the long-time behavior of form~(\ref{rho}) with the same parameters $\gamma$ and $\omega$ independent of the order of the density matrix.  On the other hand, the experimental confirmation of this relationship significantly strengthens the picture based on the notion of chaotic relaxation modes (Pollicott-Ruelle resonances).

We use the theoretical framework of Ref.~\cite{Fine-05}. The quantity called the ``signal'' is the total polarization of nuclear spins transverse to the external magnetic field. We consider the NMR response to the sequence of two rf pulses

\begin{equation}
90^{\circ}_y - \tau - X
\label{pulses}
\end{equation}
where the $90^{\circ}_y$ pulse initiates the free induction decay (FID) and, after the delay time $\tau$, pulse $X$ ``scrambles'' the time evolution of the spin system. The FID between the two pulses is to be denoted by function $F(\tau)$, and the signal at time $(t-\tau)$ after the second pulse is to be characterized by the echo response function $\tilde{F}(\tau, t)$. Time $t$ is understood to be measured from the time of the first pulse.  Most of the experimental tests of the FID-spin-echo relationships reported in Refs.\cite{Morgan-08,Sorte-11} used the solid echo pulse sequence characterized by $X=90^{\circ}_x$~\cite{Powles-62,Slichter-90}.

In the rest of this paper, we focus on the echo response $\tilde{F}(\tau, t)$, which is initiated at a time $\tau$ sufficiently long such that the FID function $F(\tau)$ has already reached the asymptotic form~\eqref{ltform}.  This regime was considered in Ref.~\cite{Fine-05},  but there the main focus was on obtaining the envelope of the Hahn spin echo sequence $90^{\circ}_y - \tau - 180^{\circ}_x-\tau$ for heteronuclear spin systems in an inhomogeneous magnetic field, in which case the echo can be monitored only at time $\tau$ following the second pulse.   In the present paper, we assume that the magnetic field is homogeneous, and thus that the echo response to the pulse sequence (\ref{pulses}) can be monitored at any moment of time following the pulse $X$. We further assume that the shapes of the FIDs and echoes are determined by the dynamics of an isolated system of interacting nuclear spins in the Larmor rotating reference frame. The interaction Hamiltonian is assumed to be non-integrable, such as the case of the standard Hamiltonian of truncated magnetic-dipolar interaction~\cite{Slichter-90}.

When $\omega \neq 0$,  the long-time behavior of the density matrix~(\ref{rho}) consists of the sum of the two Hermitian-conjugate terms $\rho_{0, kl} e^{-(\gamma + i \omega) \tau} $ and $\rho_{0, kl}^{\dagger} e^{-(\gamma - i \omega) \tau} $. Each of these terms evolves in time in a self-similar way, in the sense that the evolution is controlled by the time-independent matrix $\rho_{0,kl}$ or $\rho^{\dagger}_{0,kl}$, while the time evolution of the entire density matrix $\rho_{kl}(t)$ is reduced to rescaling each of the above terms and changing their relative phase. As a result, we can also express the long-time behavior of the FID signal as the sum of two corresponding contributions
\begin{equation}
F(\tau) = f(\tau) + f^*(\tau),
\label{Ftau}
\end{equation}
where
\begin{equation}
f(\tau) = {1 \over 2} \; a \; e^{-(\gamma + i \omega) \tau},
\label{ftau}
\end{equation}
and $a$ is a complex-valued constant.  Following pulse $X$, the new density matrix becomes

\begin{equation}
\rho_{kl}(\tau_+) = e^{-(\gamma + i \omega) \tau} \hat{U}_X \rho_{0, kl}  + e^{-(\gamma - i \omega) \tau}   \hat{U}_X  \rho_{0, kl}^{\dagger} ,
\label{r+}
\end{equation}
where $\hat{U}_X $ is the quantum operator representing the effect of pulse $X$.
As a result, we obtain
\begin{eqnarray}
 \tilde{F}(\tau, t) &=& f(\tau)  \tilde{f}(t-\tau) +  f^*(\tau)  \tilde{f}^*(t-\tau)
\nonumber
\\
 &=&|a| e^{-\gamma  \tau} \left[
\hbox{cos}(\omega \tau  - \varphi_a) \hbox{Re} \tilde{f}(t-\tau) + \hbox{sin}(\omega \tau  - \varphi_a) \hbox{Im} \tilde{f}(t-\tau) 
\right],
\label{Ftildeosc}
\end{eqnarray}
where $\tilde{f}(t-\tau)$ is the self-similar shape of the echo response associated with the first term in Eq.~(\ref{r+}), and $\varphi_a$ is the complex phase of $a$.

Equation~\eqref{Ftildeosc} implies that one can experimentally measure any two echo responses $\tilde{F}(\tau_1, t)$ and $\tilde{F}(\tau_2, t)$, such that
$\omega (\tau_2 - \tau_1)$ is not equal to a multiple of $\pi$, then extract from these two responses functions $\hbox{Re} \tilde{f}(t-\tau)$ and $\hbox{Im} \tilde{f}(t-\tau)$ and then, finally, predict $\tilde{F}(\tau, t)$ for all other $\tau$. In fact, function $\hbox{Im} \tilde{f}(t-\tau)$ can be measured directly by applying pulse $X$ at a node of the FID, where $\hbox{cos}(\omega \tau  - \varphi_a) =0 $,  while $\hbox{Re} \tilde{f}(t-\tau)$ can be measured by applying pulse $X$ in the middle between two nodes, where $\hbox{sin}(\omega \tau  - \varphi_a) =0 $. 

We can now elaborate on the long-time behavior of $\tilde{f}(t-\tau)$ in order to relate the phases of the FID beats with the phases of the echo responses.  The long-time behavior of $\tilde{f}(t-\tau)$ is expected to be of the following form\cite{Fine-05}

\begin{equation}
\tilde{f}(t-\tau) = b_1  e^{-(\gamma + i \omega) (t-\tau)}  + b_2  e^{-(\gamma - i \omega) (t-\tau)}  ,
\label{ftildelong}
\end{equation}
where $b_1$ and $b_2$ are two complex-valued constants which are not necessarily complex conjugates of each other.  (The only requiment here is that $\tilde{F}(\tau, t)$ given by Eq.~\eqref{Ftildeosc} is real.)
Substitution of Eq.~\eqref{ftildelong} into Eq.~\eqref{Ftildeosc} gives

\begin{equation}
\tilde{F}(\tau, t) = {1 \over 2} \; |a|  \; e^{-\gamma  t}  \; |C(\tau)| \; \hbox{cos}[\omega t + \varphi_C(\tau)],
\label{Ftildeosclong}
\end{equation}
where $|C(\tau)|$ and $\varphi_C(\tau)$ are the amplitude and the complex phase of the function

\begin{equation}
C(\tau) = b_1^* e^{-i \varphi_a} + b_2 e^{i(\varphi_a - 2 \omega \tau)}.
\label{C}
\end{equation}

Both $|C(\tau)|$ and $\varphi_C(\tau)$ should be independently accessible experimentally. The convenient representation of Eq.\eqref{C} to test is
\begin{equation}
|C(\tau)|^2 = |b_1|^2 + |b_2|^2 + 2 |b_1| |b_2| \hbox{cos}(2 \omega \tau - 2 \varphi_a - \varphi_{b_1} - \varphi_{b_2}),
\label{Cabs}
\end{equation}
\begin{equation}
|C(\tau)| \; \hbox{cos}\varphi_C(\tau)  = |b_1| \hbox{cos}(\varphi_a + \varphi_{b_1}) +    |b_2| \hbox{cos}(2 \omega \tau - \varphi_a -  \varphi_{b_2}),
\label{phiC}
\end{equation}
where $\varphi_{b_1}$ and $\varphi_{b_2}$ are the complex phases of constants $b_1$ and $b_2$, respectively.

The FID function $F(\tau)$ and the family of echoes $\tilde{F}(\tau, t)$ can be measured experimentally choosing $\tau$ and $t - \tau$ large enough that the long-time regime \eqref{ltform} is reached for both $F(\tau)$ and $\tilde{F}(\tau, t)$. The test of Eqs.~(\ref{Cabs}) and~\eqref{phiC} can then be carried out in the following way: 

1) The parameters $|a|$, $\varphi_a$, $\gamma$ and $\omega$ are obtained from the FID asymptotics.

2) The values of $|C(\tau)|$ and $\varphi_C(\tau)$  are obtained for each $\tau$ by fitting the tails of the echo responses $\tilde{F}(\tau, t)$ to Eq.~(\ref{Ftildeosclong}) as a function of $t$.

3) Equations (\ref{Cabs},\ref{phiC}) predict that both $|C(\tau)|^2$ and $|C(\tau)| \; \hbox{cos}\varphi_C(\tau)$ consist of two terms: a $\tau$-independent constant and a $\tau$-dependent term oscillating with frequency $2 \omega$. The observation of this behavior as a function of $\tau$ already constitutes a non-trivial test of the theory.

4) The experimental curves for $|C(\tau)|^2$ and $|C(\tau)| \; \hbox{cos}\varphi_C(\tau)$ are parameterized as follows:
\begin{equation}
|C(\tau)|^2 = B_1 + D_1 \; \hbox{cos}(2 \omega \tau + \phi_1 ),
\label{Cabsexp}
\end{equation}
\begin{equation}
|C(\tau)| \; \hbox{cos}\varphi_C(\tau)  = B_2+    D_2 \; \hbox{cos}(2 \omega \tau + \phi_2),
\label{phiCexp}
\end{equation}
where the six parameters $B_1$, $B_2$, $D_1$, $D_2$, $\phi_1$ and $\phi_2$ should be directly accessible.  The choice of the phases $\phi_1$ and $\phi_2$ are made such that the values of $D_1$  and $D_2$ are positive.  The theoretical formulas (\ref{Cabs},\ref{phiC}) depend on four real-valued parameters: $|b_1|$, $|b_2|$,
$\varphi_{b_1}$ and $\varphi_{b_2}$. Therefore, two further non-trivial tests are possible.

{\it Test 1}: One obtains $|b_2| = D_2$, and then $|b_1| = \sqrt{B_1 - |b_2|^2}$. {\it Prediction}: $D_1 = 2 |b_1| |b_2|$.

{\it Test 2}:  One obtains $ \varphi_{b_2} = - \phi_2 - \varphi_a$, and then $\varphi_{b_1} = - \phi_1 - 2 \varphi_a - \varphi_{b_2}$.  {\it Prediction}: $B_2 = |b_1| \hbox{cos}(\varphi_a + \varphi_{b_1}) = |b_1| \hbox{cos}(\phi_2 - \phi_1) $.

As a final remark, we would like to mention that, in the case of monotonic long-time decays [$\omega = 0$ in Eq.(\ref{ltform})], the treatment analogous to the one presented above predicts that the echo responses exhibit a self-similar shape with monotonic exponential long-time tails, which on a semi-logarithmic plots of the type of Fig.~\ref{fig:caf2_echoes} would fall onto the same line.

\section{\label{sec:experimental_procedures} Experiment }

This section is focused on \caf\ where we were able to initiate echoes in the long-time
regime of the FID. We also have the results from \Xe\ in solid xenon, but in this material the latest
echoes we were able to measure were not yet quite in the long-time regime of the FID. The
are also additional theoretical complications related to the polycrystalline nature of the
solid xenon\cite{Fine-12} that would make the predictions less rigorous, even if the echoes obtained
were well into the long-time portion of the FID.  As a consequence, the \Xe\ data are
included in an appendix.
 
\begin{figure}[!]
 \subfloat[]
  {\label{fig:caf2_early}\includegraphics[width=0.5 \textwidth]{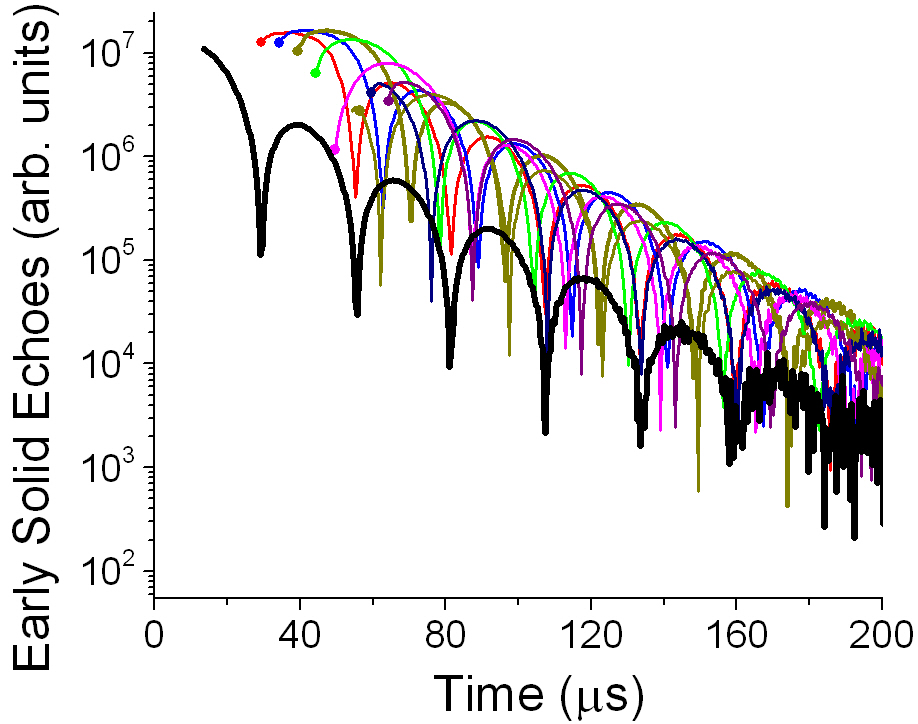}}\\       
\subfloat[]
  {\label{fig:caf2_late}\includegraphics[width=0.5\textwidth]{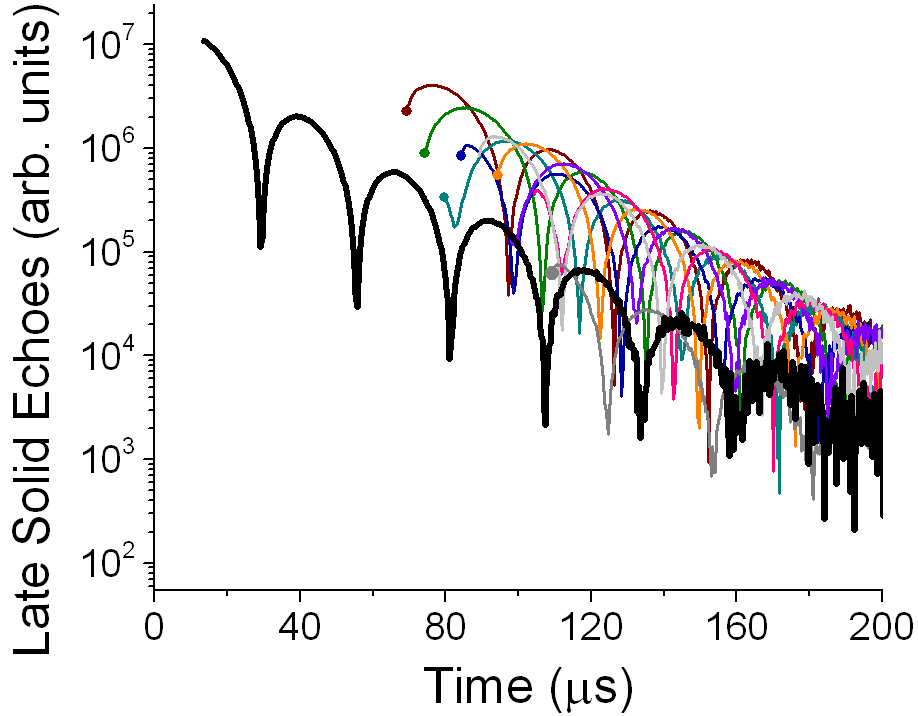}}
  \caption{(Color online).  \F\ solid echoes in \caf.  Half of the 32 acquired solid echoes are shown, split between (a) and (b) for visual clarity.  The echoes shown are acquired every $5~\mu$s from $16~\mu$s to $96~\mu$s.  The first point of each echo is indicated by a solid circle to guide the eye. }
  \label{fig:caf2_echoes}
\end{figure}

The FID and solid echoes of \F\ in \caf\ were acquired at room temperature in an external magnetic field of 2 T (\F\ Larmor frequency 83.55 MHz).  The \caf\ crystal used in our experiments was obtained from Optovac, Inc and is lightly doped with paramagnetic impurities (0.01\% Ce) to reduce T$_1$ to $\approx$ 2 sec at 2 T.  The \caf\ crystal was prepared with [100] axis along the long dimension of the cylinder.  The crystal was then held with the magnetic field along the [001] direction.  The data were acquired using the same Tecmag spectrometer using 2 $\mu$s square pulses with a receiver dead time of 13 $\mu$s.  Using these parameters, 1000 transients were averaged with a repetition time of 10 seconds for each experiment to enhance the signal-to-noise ratio. 32 solid echoes were acquired, one every $2.5~\mu$s from $16~\mu$s to $96~\mu$s.  At $60~\mu$s, fits to the FID show that it has entered the long-time regime described by Eq.~\eqref{ltform}; therefore, echoes generated after time $t= 60~\mu$s meet the criteria for testing the predictions made in Section~\ref{sec:theory}.  The family of representative echoes is shown in Fig.~\ref{fig:caf2_echoes}. 

Determination of the amplitude $|a|$, decay coefficient $\gamma$, beat frequency $\omega$, and complex phase $\varphi_a$ of the FIDs in CaF$_2$ was made by fits to Eq.~\eqref{ltform} (see Table~\ref{table1}).  The amplitude $|C(\tau)|$ and complex phase $\varphi_C$ of each echo signal were determined by fitting each solid echo signal to Eq.~\eqref{Ftildeosclong} with $\gamma$, $\omega$, $|a|$, and $\varphi_a$ fixed to the values obtained from the fit to the FID. 

\begin{table}
\begin{tabular}{ | l | l |}
\hline
 FID Parameters & \hspace{1cm} Solid Echo Parameters \\
  \hline
   $|a|$ = $9,380,000  \pm 45,000 $ & $B_1$ = $18.3 \pm 0.2$ \ \  \ $B_2 = 0.02 \pm 0.3$ \\
 $\varphi_a = 1.921 \pm 0.006$  &  $D_1$ = $2.1 \pm 0.3$ \ \ \  \ $D_2 = 4.2 \pm 0.4$  \\
 $\gamma = 0.0414 \pm .0008\ \mu$s$^{-1}$ &  $\phi_1$ = $-1.5 \pm 0.1$ \ \  \ $\phi_2= 0.4 \pm 0.1$\\
  $\omega = 0.120\pm .007\ \mu$s$^{-1}$ & \\ 
  \hline 
\end{tabular}
\caption{Long-time FID and echo fit parameters for \caf.  The FID parameters are obtained by fitting the FID to Eq.~\eqref{ltform}.  The echo parameters are obtained by fitting the amplitudes and phases of the measured solid echoes to Eqs.\eqref{Cabsexp} and \eqref{phiCexp}. }
\label{table1}
\end{table}

\begin{figure}[!]
 \subfloat[]
  {\label{fig:caf2_amps}\includegraphics[width=0.5 \textwidth]{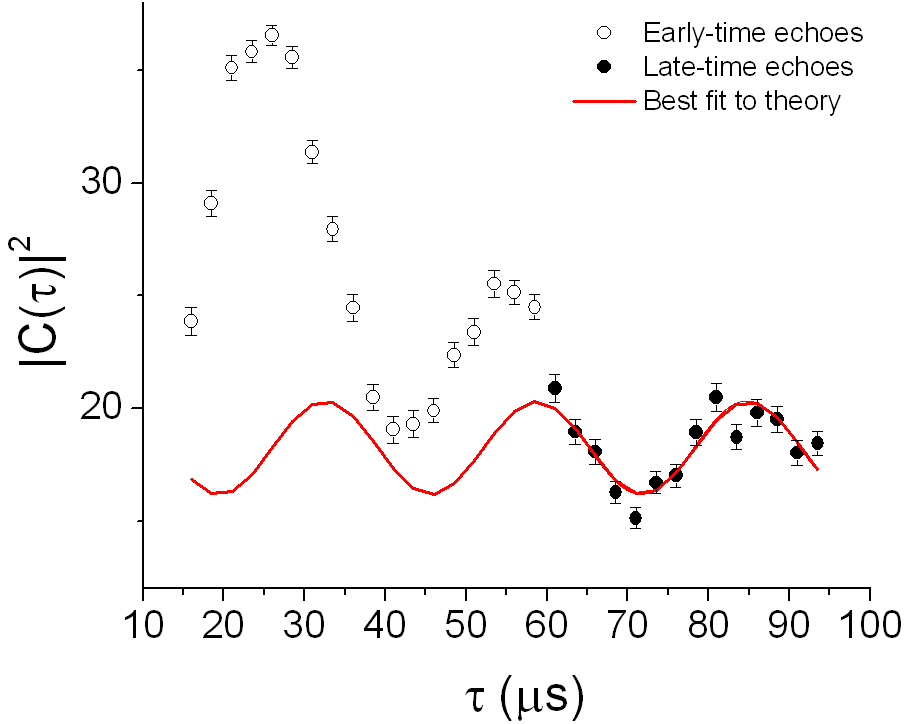}}       \\
\subfloat[]
  {\label{fig:caf2_phases}\includegraphics[width=0.5\textwidth]{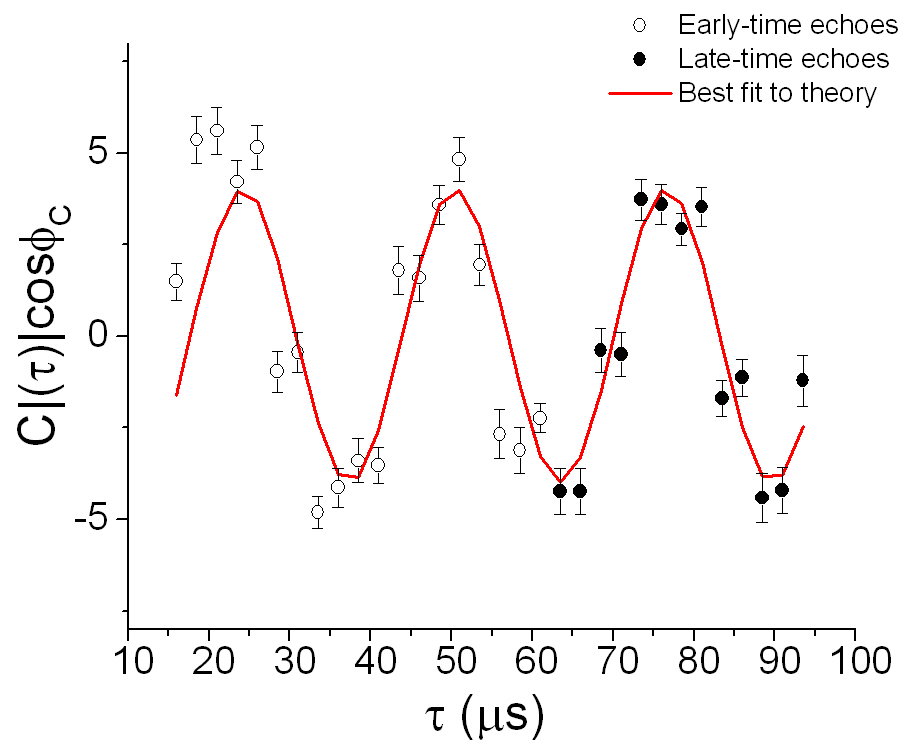}}
  \caption{ (Color online) (a) Amplitudes and (b) phases of \F\ solid echoes in \caf\ as a function of inter-pulse delay time $\tau$.  Open circles represent echoes generated in the early-time of the FID, while solid circles represent echoes generated in the long-time of the FID.  The solid line (red) is the best fit of the long-time data to Eq.~\eqref{Cabsexp} for (a) or to Eq.~\eqref{phiCexp} for (b).}
  \label{fig:caf2_amp_phase}
\end{figure}

In Fig.~\ref{fig:caf2_amp_phase} we plot $|C(\tau)|^2$ and $|C(\tau)| \cos\phi_C$  for the solid echoes of \F\  in \caf.  The solid lines are the fits to either Eq.~\eqref{Cabsexp} or Eq.~\eqref{phiCexp}, from which the parameters in Table~\ref{table1} were obtained.  Figure~\ref{fig:caf2_amp_phase} shows the results for echoes initiated in both the early-time and the long-time regimes of the FID in order  to illustrate the approach to the long-time behavior described by Eqs.(\ref{Cabsexp},\ref{phiCexp}). 
 
Tests 1 and 2 formulated at the end of Section~\ref{sec:theory}  are then carried out.  The predicted and the measured values of parameters $D_1$ and $B_2$ are listed in Table~\ref{table2}.  We find that in each test the predicted and the measured values agree with each other within the experimental uncertainties.

\begin{table}[!]
\begin{tabular}{| c | c | c |}
  \hline
 {\it Test 1}  & {\it Test 2} \\
\hline
  $D_{1, predicted} = 2.3 \pm 0.3 $ & $B_{2, predicted} = -0.01 \pm 0.07 $ \\
  $D_{1, measured} = 2.1 \pm 0.3\ \ $ & $B_{2, measured} = 0.00 \pm 0.3 $ \\
  \hline 
\end{tabular}
\caption{Two tests of the theoretical predictions formulated at the end of Section~\ref{sec:theory}.}
\label{table2}
\end{table}

\begin{figure}[tbp]
\includegraphics[width=0.5  \textwidth]{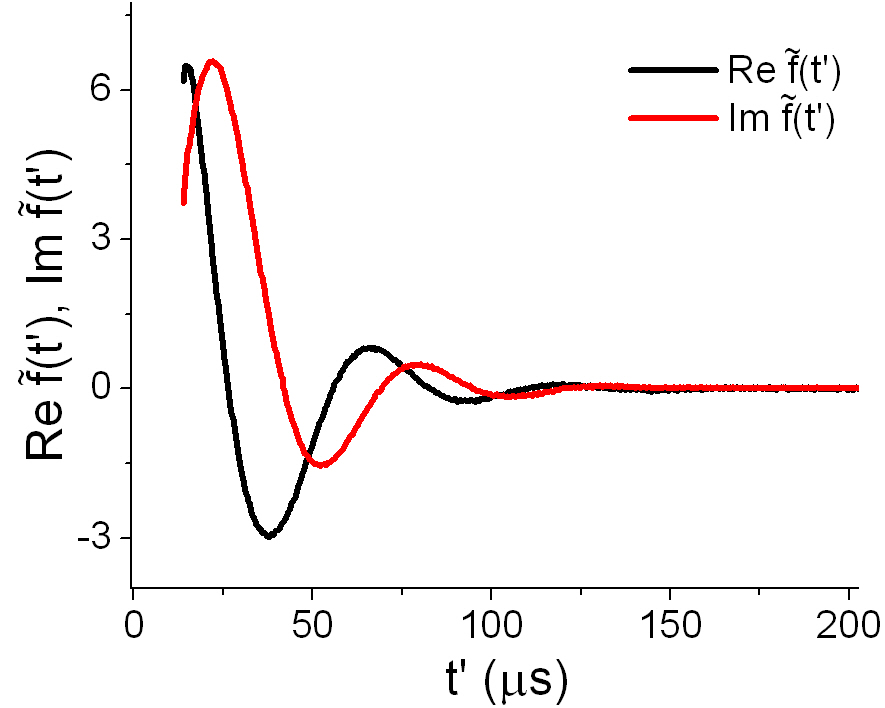} 
\caption{ (Color online) The shape functions Re$\tilde{f}(t')$ and Im$\tilde{f}(t')$ obtained from the linear system of equations~(\ref{eq1},\ref{eq2}).}
\label{fig:re_im}
\end{figure}

Finally,  Section~\ref{sec:theory} contains a more general prediction: that all possible shapes of echo responses including both the initial and the long-time behavior can be obtained from two basic functions Re$\tilde{f}(t-\tau)$ and Im$\tilde{f}(t-\tau)$ as described by Eq.~\eqref{Ftildeosc}.  To determine these functions, we chose two measured echo responses $\tilde{F}(\tau_1,t)$ and $\tilde{F}(\tau_2,t)$  initiated at times $\tau_1=81.0\ \mu s$ and $\tau_2=93.5\ \mu s$, respectively.  According to Eq.~\eqref{Ftildeosc} 
\begin{eqnarray}
\tilde{F}(\tau_1,t'+\tau_1) = A_1 \hbox{Re}\tilde{f}(t') + B_1 \hbox{Im}\tilde{f}(t') \label{eq1} \\ 
\tilde{F}(\tau_2,t'+\tau_2) = A_2 \hbox{Re}\tilde{f}(t') + B_2 \hbox{Im}\tilde{f}(t') 
\label{eq2}
\end{eqnarray}
where in Eq.(\ref{eq1}) $t'=t-\tau_1$, $A_1 = F(\tau_1)$ and  $B_1 = F(\tau_1) \tan (\omega \tau_1 - \varphi_a)$, while in Eq.(\ref{eq1}) $t'=t-\tau_2$, $A_1 = F(\tau_2)$ and  $B_1 = F(\tau_2) \tan (\omega \tau_2 - \varphi_a)$. Here $F(\tau_1)$ and $F(\tau_2)$ are the measured values of the FID at times $\tau_1$ and $\tau_2$, respectively.    Now we express functions $\hbox{Re}\tilde{f}(t')$ and $\hbox{Im}\tilde{f}(t')$ in terms of the measured functions $\tilde{F}(\tau_1,t'+\tau_1) $ and $\tilde{F}(\tau_2,t'+\tau_2) $  by solving the system of linear equations (\ref{eq1},\ref{eq2}) and then substitute $\hbox{Re}\tilde{f}(t - \tau)$ and $\hbox{Im}\tilde{f}(t-\tau)$  back to Eq.(\ref{Ftildeosc}) to predict other  echo responses initiated in the long-time FID regime. The functions $\hbox{Re}\tilde{f}(t')$ and $\hbox{Im}\tilde{f}(t')$ are plotted in Fig.~\ref{fig:re_im}.  Since for solid echoes $\tilde{F}(\tau,\tau) = F(\tau)$, Eq.(\ref{Ftildeosc}) implies that $\hbox{Re}\tilde{f}(0)=1$ and $\hbox{Im}\tilde{f}(0)=0$; however, due to the finite ($13\ \mu$s) recovery time of our spectrometer, we were unable to measure these functions back to time $t'=0$.  

\begin{figure}[tbp]
\includegraphics[width=0.5  \textwidth]{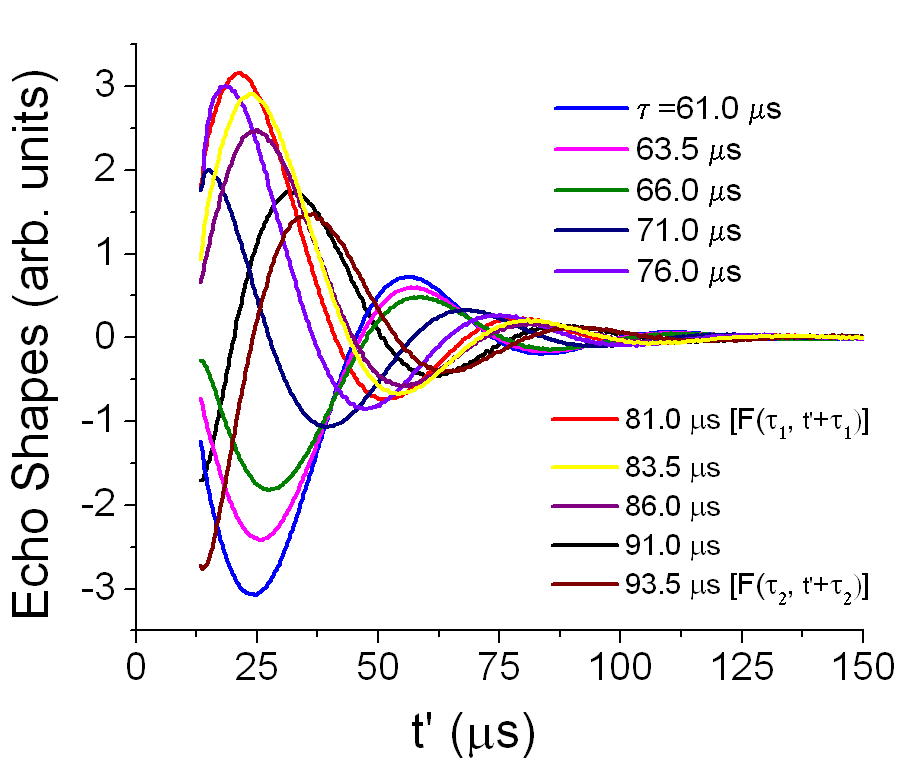} 
\caption{ (Color online) \F\ solid echoes in \caf\ labeled by their interpulse delay times $\tau$.  The quantity plotted is $e^{\gamma \tau} \tilde{F}(\tau, t'+\tau)$.  The $81.0~\mu$s and $93.5~\mu$s echoes represent the echoes used to obtain the shape functions Re$\tilde{f}(t')$ and Im$\tilde{f}(t')$.  }
\label{fig:combined}
\end{figure}

\begin{figure}[tbp]
\includegraphics[width=0.5  \textwidth]{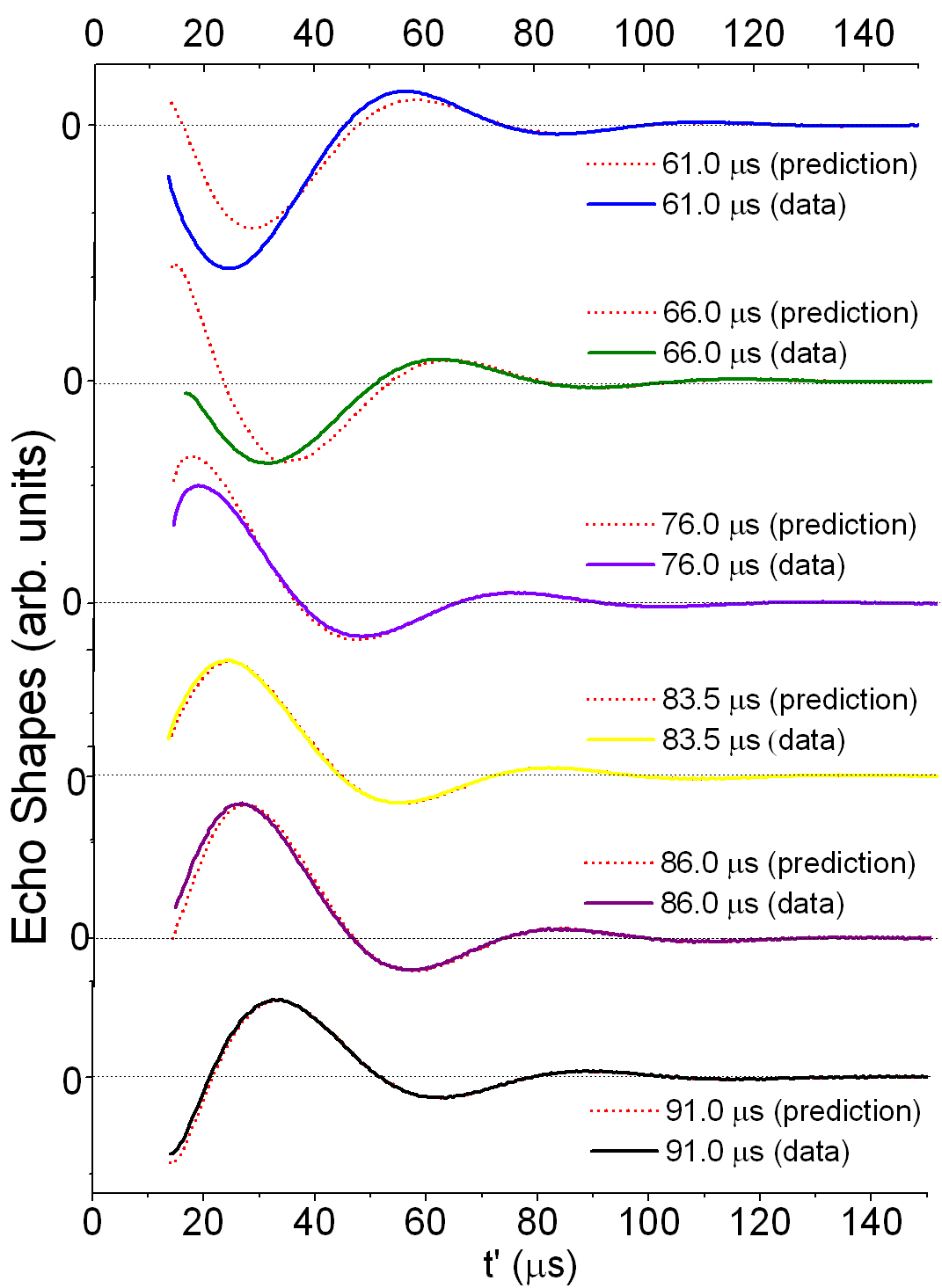} 
\caption{ (Color online) \F\ solid echoes in \caf\ (solid lines).  The red (broken) lines show the theoretical echo shapes obtained using the Re$\tilde{f}(t')$ and Im$\tilde{f}(t')$ shape functions.  The quantity plotted is $e^{\gamma \tau} \tilde{F}(\tau, t'+\tau)$.  The values of $\tau$ are indicated in the plot legend.}
\label{fig:all_echoes}
\end{figure}

In Fig.~\ref{fig:combined}, we show several of the measured echoes in \caf.  Each echo $\tilde{F}(\tau, t'+\tau)$ in Fig.~\ref{fig:combined} has been multiplied by a factor $e^{\gamma \tau}$ to correct for the exponential decay of the FID at their respective  initial time values.  Labeled in the plot legend are the two echoes $\tilde{F}(\tau_1, t'+\tau_1)$ and $\tilde{F}(\tau_2, t'+\tau_2)$ used in determining the shape functions $\hbox{Re}\tilde{f}(t')$ and $\hbox{Im}\tilde{f}(t')$.  In Fig.~\ref{fig:all_echoes}, the measured echo responses initiated at $\tau = 61, 66, 76, 83.5, 86$, and $91\ \mu$s are compared with the predicted ones obtained by substituting $\hbox{Re}\tilde{f}(t')$ and $\hbox{Im}\tilde{f}(t')$ into Eq.(\ref{Ftildeosc}).  We observe in Fig.~\ref{fig:all_echoes} that the agreement of the long-time behavior between the predicted and the measured echo responses is very good for all echoes, which is consistent with the results presented in Table~\ref{table2} and in Fig.~\ref{fig:caf2_amp_phase}.  The initial behavior of the early echo responses ($\tau = 61\ \mu$s, $66\ \mu$s, and $76\ \mu$s) exhibits some discrepancies between the predicted and the measured behavior. However, it is clear that the predicted behavior still captures the evolution of the measured echo shapes in a satisfactory way.  In the later echoes ($\tau = 83.51\ \mu$s, $86\ \mu$s, and $91\ \mu$s), this initial discrepancy no longer appears and the entire echo shape is found to agree with the predicted shape.

A possible reason for the above initial discrepancy in the early echoes may be the presence of the chaotic eigenmodes that decay faster than the slowest mode controlling the long-time FID behavior (\ref{ltform}), but still not fast enough to completely disappear by time $\tau$ when the echo is initiated. The existence of a well-separated second slowest eigenmode was demonstrated by the recent experiment of Meier et al.\cite{Meier-11} on a \caf\ crystal for the same orientation with respect to the magnetic field.  This second eigenmode disappears below the noise level on the timescale of $60\ \mu$s.     
These additional eigenmodes are probably more pronounced in the higher-order spin correlations, because these correlations develop in the course of the FID evolution only after an initial  time delay with respect to the beginning of the FID\cite{Cho-05}.  Therefore, the behavior of many-spin density matrices should approach the long-time form (\ref{rho}) also with some delay with respect to the time when the FID starts exhibiting the universal long-time form (\ref{ltform}).  The same additional eigenmode may be controlling the approach of $|C(\tau)|^2$ to the predicted asymptotic behavior in Fig.~\ref{fig:caf2_amps}.

\section{\label{sec:conclusion} Summary and conclusions}
In this paper, we investigated the properties of spin echoes initiated in the regime where the nuclear FID has reached the universal exponentially damped oscillatory behavior. Using the theoretical framework motivated by the notion of microscopic chaos, we predicted how the shapes of the echo responses depend on the phase of the FID oscillations at the time of the echo pulse, and, in particular, obtained the phase relationships between the long-time oscillation of the FIDs and the echoes. We have further conducted several experimental tests of the above predictions for FIDs and solid echoes in CaF$_2$ and solid xenon, and obtained results in good overall agreement with the theoretical expectations. The long-time phase relationships between the FID and the echoes were confirmed particularly well.  This good agreement amounts to an indication that the long-time behavior of the higher-order spin density matrices has the form given by Eq.~\eqref{rho} with the same values of $\gamma$ and $\omega$ as the original FID.  Such a behavior is expected for a relaxational eigenmode of the time-evolution operator in a chaotic system. 

While the fundamental difficulties in defining the notion of microscopic chaos still remain, the present paper demonstrates that the approach of Refs.~\cite{Fine-04,Fine-05} based on making parallels with relaxational eigenmodes in classical chaotic systems continues to generate successful quantitative predictions.  These predictions were made in a regime not accessible by controllable first principles calculations. We are not aware of any other approach that would reproduce the same predictions under conditions that the quantities of interest (nuclear spin decays) evolve on the fastest microscopic time scale of the system. 

\begin{acknowledgments}

This work was supported by NSF grant PHY-0855482.

\end{acknowledgments}
\appendix


\section{Measurements in Solid Xenon}

We have carried out measurements on \Xe\ in solid xenon similar to those reported in Sec.~\ref{sec:experimental_procedures} on \caf, but due to experimental constraints described below, we were not able to generate solid echoes in the long-time regime of the FID [i.e. where the FID is well-described by Eq.~\eqref{ltform}] that themselves had enough signal-to-noise ratio to be measured into their respective long-time regimes.  However, as the echoes acquired with the longest interpulse delays show an approach to the predicted forms, we present the data even though, strictly speaking, they were not generated in a regime wherein the predictions hold.    

\begin{figure}[htbp]
\begin{center}
\includegraphics[width=0.5\textwidth]{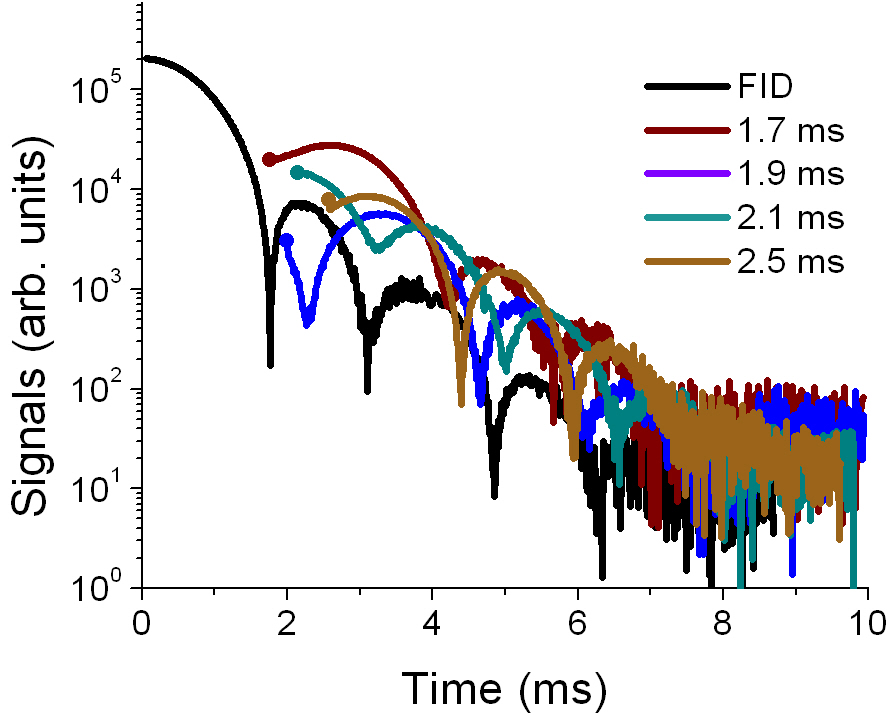}  
\caption{(Color online).  \Xe\ solid echoes in solid xenon.  The latest 4 echoes acquired are shown.  The first point of each echo is indicated by a solid circle to guide the eye.}
\label{fig:xe_echoes}
\end{center}
\end{figure}

For this appendix we use the FID and solid echoes which are presented in Ref.~\cite{Sorte-11}.  Polycrystalline xenon samples were produced in a magnetic field of 2 T (\Xe\ Larmor frequency 24.56 MHz) using the methods described in Ref.~\cite{Sorte-11}.  Ten solid echoes were acquired approximately 0.2~ms apart from 0.4~ms to 2.5~ms.  A fit of Eq.~\eqref{ltform} to the FID show that it does not enter the long-time regime described by Eq.~\eqref{ltform} until after $t=2.5$~ms.  As the latest echo was acquired with an interpulse delay time of $2.5 $ ms, no echoes were acquired in the long-time regime of the FID.  The xenon FID and solid echoes are shown in Fig.~\ref{fig:xe_echoes}.

The amplitude $|a|$, decay coefficient $\gamma$, beat frequency $\omega$, and complex phase $\varphi_a$ of the \Xe\ FID were determined by fits to Eq.~\eqref{ltform} (see Table~\ref{table3}).  The amplitude $|C(\tau)|$ and complex phase $\varphi_C$ of each echo signal were determined by fitting each solid echo signal to Eq.~\eqref{Ftildeosclong} with $\gamma$, $\omega$, $|a|$, and $\varphi_a$ fixed to the values obtained from the fit to the FID. 

\begin{table}[b]
\begin{tabular}{| l | l |}
\hline
FID Parameters & \hspace{1cm} Solid Echo Parameters \\
  \hline
  $|a|$ = $151,800 \pm 1,600 $ & $B_1 = 260 \pm 30$ \ \ \ \ $B_2= 0.1\pm 3$ \\
 $\varphi_a = -1.254 \pm 0.006$ & $D_1 = 290 \pm 40$ \ \ \ \ $D_2=12\pm 3$ \\
$\gamma = 1.251 \pm 0.005$ ms$^{-1}$ & $\phi_1 = 0.71 \pm 0.05$ \ \ \ \ $\phi_2=2.4\pm 0.5$ \\
  $\omega = 2.10 \pm 0.01$ ms$^{-1}$ & \\ 
  \hline 
\end{tabular}
\caption{Long-time FID and echo fit parameters for solid xenon.  The FID parameters are obtained by fitting the FID to Eq.~\eqref{ltform}.  The echo parameters are obtained by fitting the amplitudes and phases of the measured solid echoes to Eqs.\eqref{Cabsexp} and \eqref{phiCexp}. }
\label{table3}
\end{table}

\begin{figure}[!]
 \subfloat[]
  {\label{fig:xe_amps}\includegraphics[width=0.45\textwidth]{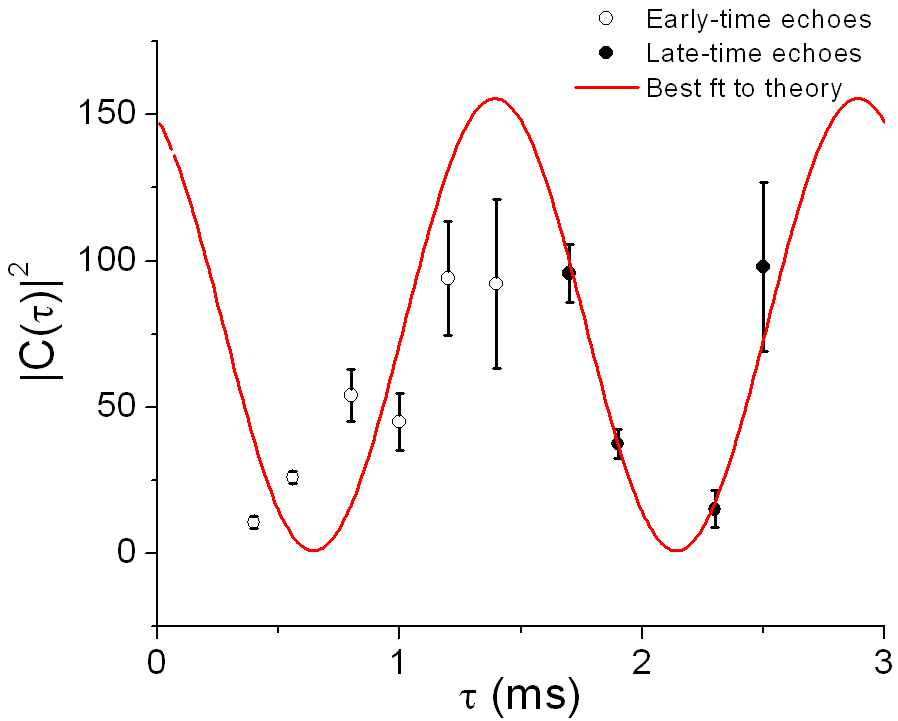}}    \\       
\subfloat[]
  {\label{fig:xe_phases}\includegraphics[width=0.45\textwidth]{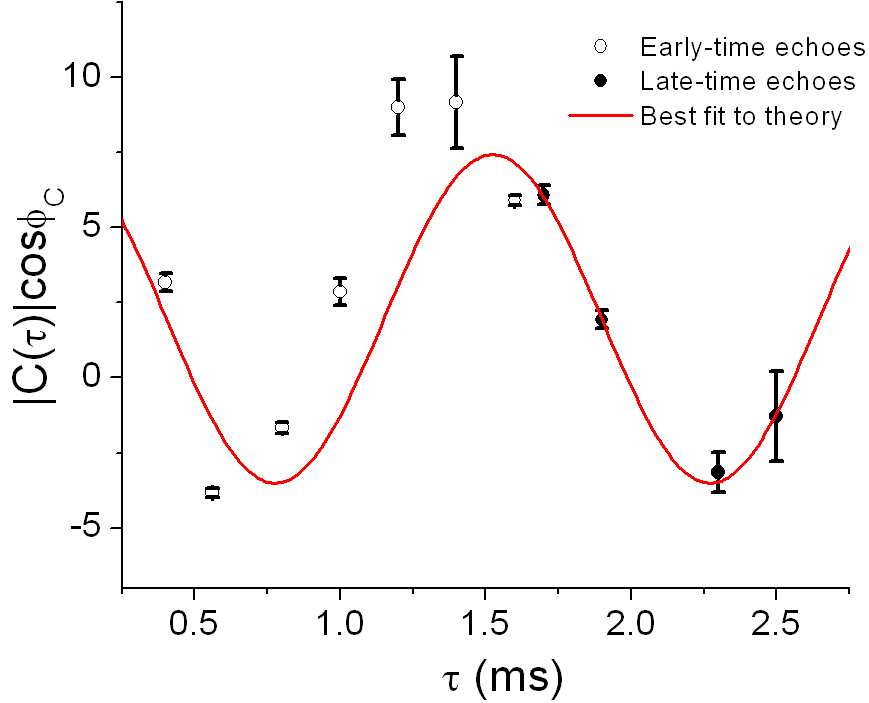}}
  \caption{ (Color online) (a) Amplitudes and (b) phases of $^{129}$Xe solid echoes in solid xenon as a function of inter-pulse delay time $\tau$.  Open circles represent echoes generated in the early-time of the FID, while solid circles represent echoes generated in the long-time of the FID.  The solid line (red) is the best fit of the long-time data to Eq.~\eqref{Cabsexp} for (a) or to Eq.~\eqref{phiCexp} for (b). }
  \label{fig:xe_amp_phase}
\end{figure}

In Fig.~\ref{fig:xe_amp_phase}, we plot $|C(\tau)|^2$ and $|C(\tau)| \cos\phi_C$  for the measured solid echoes.  The solid lines are the fits to either Eq.~\eqref{Cabsexp} or Eq.~\eqref{phiCexp}, from which the parameters in Table~\ref{table3} were obtained.  Figure~\ref{fig:xe_amp_phase} shows the results for echoes initiated in both the early-time and the long-time regimes of the FID.  We observe that the latest echoes begin to approach the behavior predicted by Eqs.~\eqref{Cabsexp} and~\eqref{phiCexp}.  These echoes are labeled ``late-time echoes" even thought they are not actually in the late-time region as is the case in the \caf.

Tests 1 and 2 formulated at the end of Section~\ref{sec:theory}  are then carried out.  The predicted and the measured values of parameters $D_1$ and $B_2$ are listed in Table~\ref{table4}.  In each test, the predicted and the measured values agree with each other within the experimental uncertainties.

\begin{table}[!]
\begin{tabular}{| c | c |}
  \hline
 {\it Test 1}  & {\it Test 2} \\
\hline
 $D_{1, predicted} = 250 \pm 60 $ & $B_{2, predicted} = -1.0 \pm 0.3 $ \\
 $D_{1, measured} = 290 \pm 40 $ & $B_{2, measured} = 0 \pm 3 $ \\
  \hline
\end{tabular}
\caption{Two tests of the theoretical predictions formulated at the end of Section~\ref{sec:theory}.}
\label{table4}
\end{table}

\begin{figure}[!]      
\subfloat[]
  {\label{xe_scaled_echoes}\includegraphics[width=0.5\textwidth]{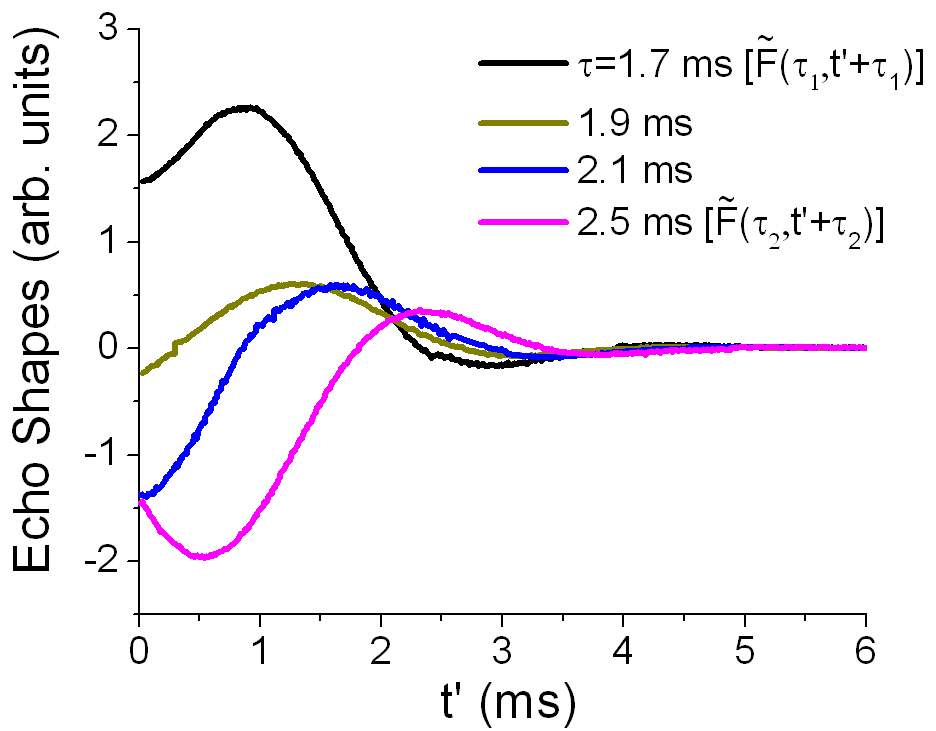}}\\
   \subfloat[]
  {\label{fig:xe_re_im}\includegraphics[width=0.5 \textwidth]{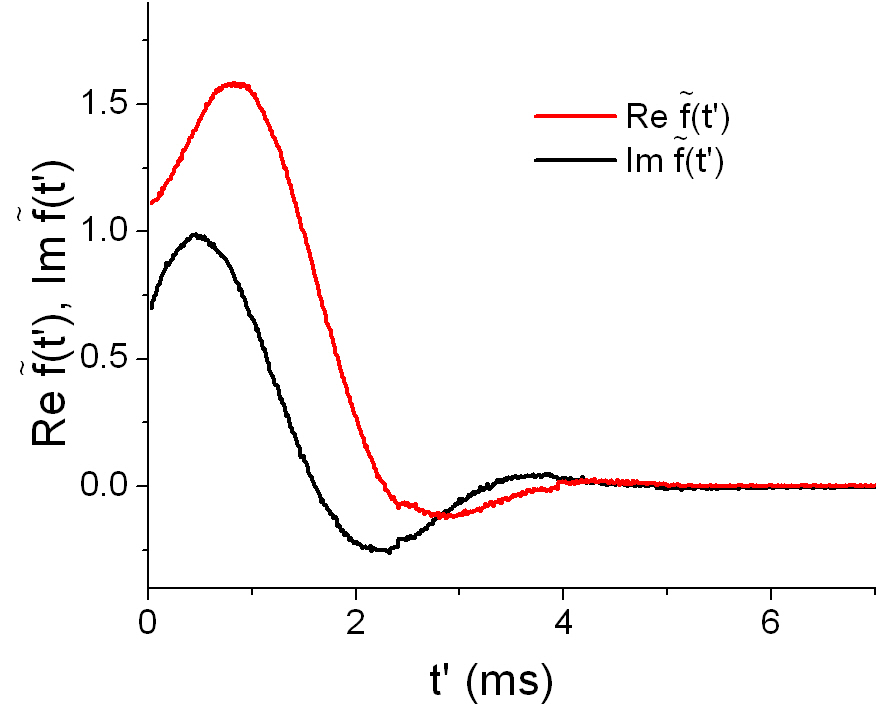}} 
  \caption{ (Color online) (a) \Xe\ solid echoes in solid xenon labeled by their interpulse delay times $\tau$.  The quantity plotted is $e^{\gamma \tau} \tilde{F}(\tau, t'+\tau)$.  The 1.7~ms and 2.5~ms echoes represent the echoes used to obtain the shape functions Re$\tilde{f}(t')$ and Im$\tilde{f}(t')$. (b) The functions Re$\tilde{f}(t')$ and Im$\tilde{f}(t')$ obtained from the linear system of equations~(\ref{eq1},\ref{eq2}). }
  \label{fig:xe_plots}
\end{figure}

Finally, we compare the obtained echoes with the predicted shape functions.  We first obtain the shape functions $\hbox{Re}\tilde{f}(t')$ and $\hbox{Im}\tilde{f}(t')$ as described in Sec.~\ref{sec:experimental_procedures}.  In Fig.~\ref{xe_scaled_echoes}, we show the measured echoes in \Xe.  Each echo $\tilde{F}(\tau, t'+\tau)$ in the figure has been multiplied by a factor $e^{\gamma \tau}$ to correct for the exponential decay of the FID at their respective  initial time values.  The two echoes $\tilde{F}(\tau_1, t'+\tau_1)$ and $\tilde{F}(\tau_2, t'+\tau_2)$ used for determining the functions $\hbox{Re}\tilde{f}(t')$ and $\hbox{Im}\tilde{f}(t')$ are shown in Fig.~\ref{xe_scaled_echoes}.  The functions $\hbox{Re}\tilde{f}(t')$ and $\hbox{Im}\tilde{f}(t')$ are shown in Fig.~\ref{fig:xe_re_im}. 

\begin{figure}[htbp]
\begin{center}
\includegraphics[width=0.47\textwidth]{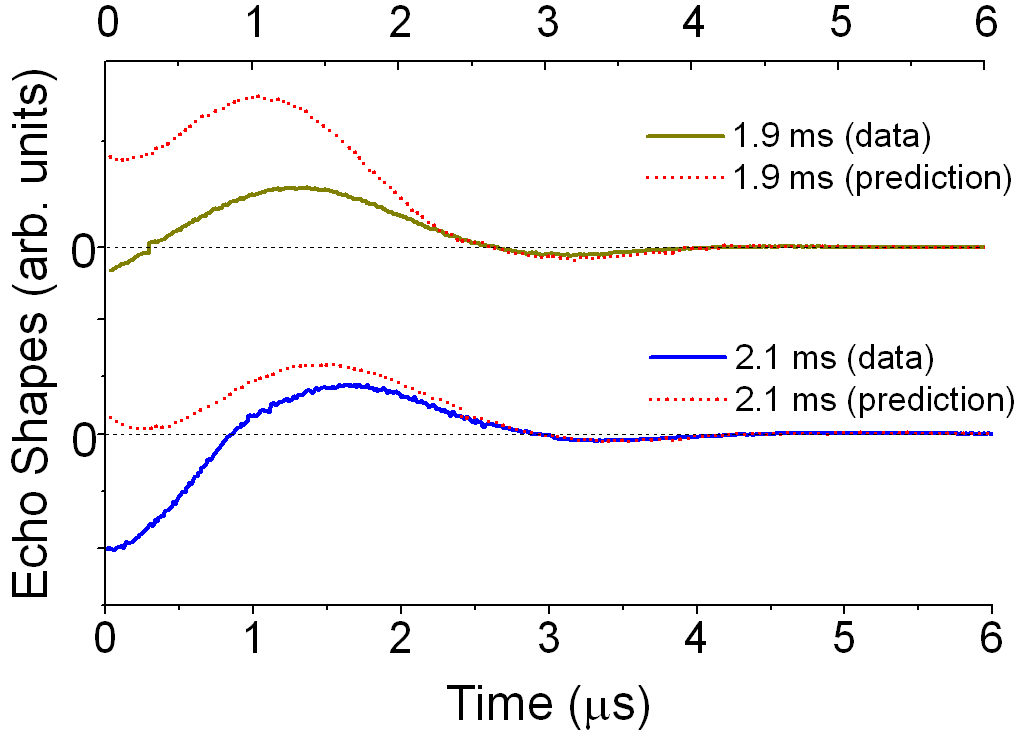}  
\caption{(Color online) \Xe\ solid echoes in solid xenon (solid lines).  The red (broken) lines show the theoretical echo shapes obtained using the Re$\tilde{f}(t')$ and Im$\tilde{f}(t')$ shape functions.  The quantity plotted is $e^{\gamma \tau} \tilde{F}(\tau, t'+\tau)$.  The values of $\tau$ are indicated in the plot legend.}
\label{xe_pred}
\end{center}
\end{figure}

In Fig.~\ref{xe_pred}, the remaining two late-time echoes are compared with the predicted ones obtained by substituting $\hbox{Re}\tilde{f}(t')$ and $\hbox{Im}\tilde{f}(t')$ into Eq.(\ref{Ftildeosc}).  We emphasize that the theoretical prediction of the echo shapes is expected to hold only for echoes initiated \emph{after} 2.5~ms in this material.  Therefore, the discrepancy in the early parts of the echo shapes is expected for the reasons discussed in Sec.~\ref{sec:experimental_procedures} in relation to the early-time echo shapes in \caf.

We finally mention a possible additional complication in solid xenon associated with the fact that our xenon samples are not single crystals but rather polycrystallites.  Our very recent theoretical study\cite{Fine-12} indicates that the observed long-time FID behavior of the polycrystalline solid xenon probably represents a typical long-time behavior of the individual crystallites contributing to the polycrystallite average, but the true asymptotic FID behavior of the entire polycrystallite is expected to appear only at times beyond the range accessible in our experiments. This asymptotic FID behavior  should be controlled by the small fraction of the constituent crystallites with the slowest exponential decay constants $\gamma$.


\end{document}